\documentclass[aps,prl,floats,twocolumn,showpacs]{revtex4}
\begin{document}
\title{Spin-pairing instabilities at the coincidence of two Landau
  levels}
\author{Eros Mariani$^{1}$, Nicodemo Magnoli$^{2a}$, Franco
  Napoli$^{2b}$, Maura Sassetti$^{2b}$ and Bernhard Kramer$^{1}$
  \vspace{1mm}}
\affiliation{$^{1}$I. Institut f\"ur Theoretische Physik,
  Universit\"at Hamburg,
  Jungiusstra\ss{}e 9, D-20355 Hamburg, Germany \\
  $^{2}$Dipartimento di Fisica, INFN,
  Universit\`{a} di Genova, Via
  Dodecaneso 33, I-16146 Genova, Italy\\
$^{3}$Dipartimento di Fisica, INFM,
  Universit\`{a} di Genova, Via
  Dodecaneso 33, I-16146 Genova, Italy  
\vspace{3mm}} 
\date{May 30, 2002}
\begin{abstract}The effect of interactions near the 
  coincidence of two Landau levels with opposite spins at filling factor 1/2
  is investigated. By mapping to Composite Fermions it is shown that the
  fluctuations of the gauge field induces an effective attractive Fermion
  interaction. This can lead to a spin-singlet ground state that is separated
  from the excited states by a gap. The magnitude of the gap is evaluated. The
  results are consistent with the recently observed half-polarized states in
  the FQHE at a fixed filling factor. It is suggested that similar anomalies
  exist for other spin configurations in degenerate spin-up and spin-down
  Landau levels. An experiment for testing the spin-singlet state is proposed.
\end{abstract}
\maketitle
It is well established that in two dimensions (2D) the interacting
electrons in a half-filled Landau level (LL) can be mapped via the
Chern-Simons transformation to a weakly interacting Fermi liquid of
Composite Fermions (CF) \cite{Jain89,HLR93,Books}. This singular gauge
transformation is equivalent to attaching two flux quanta to each
electron. The CF feel an effective magnetic field that is smaller than
the external one. It vanishes, on the average, at half filling. By
introducing the CF, one can understand the principal features of the
incompressible states in the fractional quantum Hall effect (FQHE) as
an integer quantum Hall effect of CFs. The CF also help clarifying the
nature and the properties of the compressible phases at other
even-denominator filling factors in the first LL \cite{HLR93}. The
theoretical expectations concerning the transport properties of the
latter states have been confirmed in surface-acoustic waves
experiments. In transport measurements on anti-dot lattices near half
filling, commensurability oscillations of CFs were observed. This
strongly suggests that CFs can be viewed as real objects which, in
certain situations, behave almost as classical point-like particles
\cite{willett}. This view is strongly supported by the recent
report of CF cyclotron resonance \cite{k02}.

\vspace{-3mm}
Via radiative recombination of electrons in the inversion layer of
high-electron mobility AlGaAs-heterostructures with holes bound to acceptors
in the delta-doping region, the spin polarizations of several FQHE-ground
states at fixed filling factors have been measured \cite{Kukushkin} as a
function of the ratio between Zeeman and the Coulomb energies, $\xi\equiv
E_{\rm Z}/E_{\rm C}$. Crossovers between FQHE-ground states with different
spin polarizations have been detected. Upon varying $\xi$, the polarization of
the ground state remains constant within large intervals until a certain
critical value $\xi_{\rm c}$ is approached. Then, the system is transferred
into a stable, differently polarized ground state. The experimental data have
been found to be consistent with the model of non-interacting CFs with spin,
and with an effective mass that scales with Coulomb interaction ($\propto
\sqrt{B}$). The broad plateaus in the spin-polarization are then due to the
occupation of a fixed number of spin split LL of the CF (CFLL). The crossovers
occur when intersections of CFLLs coincide with the chemical potential.

\vspace{-1mm}
Most strikingly, the experimentally determined spin polarizations, when
extrapolated to zero temperature, show additional plateaus for magnetic flux
densities near the centers of the crossovers. The corresponding polarizations
are within the experimental uncertainty almost exactly intermediate between
those in the neighboring broad plateaus. This indicates additional features
beyond the non-interacting CF model and could be signature of a new collective
state since one can expect that if two CFLLs are degenerate, residual
interactions become very important and cannot be treated perturbatively. In
these experiments, the CFLLs with different spins have been tuned to
degeneracy via the dependence of the effective mass of the CFs on the magnetic
field which was aligned perpendicular to the 2D system.

By measuring via NMR the spin polarization at filling factor
$\nu=2/3$, a remarkably abrupt transition from a fully polarized state
to a state with polarization 3/4 has been found when decreasing $\xi$
\cite{f01}. This has been associated with a first order quantum phase
transition. For $\nu> 2/3$, strong depolarization has been observed
that is associated with two spin flips per additional flux quantum. In
these measurements, $\xi$ has been tuned via tilting the magnetic
field such that the electron system is also subject to an in-plane
component of the magnetic field.

The nature of collective states under such conditions has been addressed
in several works. For instance, several spin polarization instabilities has
been found assuming the tilted field geometry for LLs
\cite{Giuliani85,Yarlagadda91}. Directly related to the above optical
measurements, a non-translationally invariant charge-density-wave-state of CFs
has been proposed on the basis of restricted Hartree-Fock calculations
\cite{Murthy00}. From exact diagonalizations of few interacting Fermions, a
liquid of non-symmetric excitons has been suggested \cite{a01}.

In this paper, it is proposed that a condensate of spin-singlet pairs of CF,
could account for some of the anomalies observed near degeneracy. We
investigate the coincidence of two LLs with opposite spins using the
Chern-Simons gauge transformation and considering the effective interactions
induced by the gauge field. The two intersecting LL can then be imagined as
two degenerate CF-Fermi seas with opposite spins coupled by interactions. Our
main result is that the fluctuations of the gauge field generate an effective
interaction between CFs with opposite spins and momenta which is attractive in
the limit of small frequency and long wavelengths. This indicates an
instability towards a condensate of spin-singlet pairs. The gap between the
ground state and the energetically lowest excited state is evaluated by
solving the Eliashberg equation, similar to earlier work
\cite{Khves93,Bonesteel93}.  We suggest that the half-polarized states of CFs
observed in the experiments mentioned above can be understood by generalizing
our model to pairing of second-generation CFs. We provide a quantitative
estimate of the gap which is found to be roughly consistent with the
experimental data. Similar to the previously discussed pairing instability in
well-separated double layer QHE-systems \cite{Bonesteel96}, the condensate of
CFs should show macroscopic quantum effects that could be experimentally
addressed.

We consider two half-filled, degenerate LL with opposite spin. Using
the Chern-Simons transformation we obtain CFs with spin that form 2D
Fermi seas with a Fermi wave number $k_{\rm F}=\sqrt{2\pi \rho}$
($\rho$ total average electron number density) \cite{HLR93}. This
system is formally analogous to a double-layer quantum Hall system
with $\nu=1/2$ in each layer. The layer index is here the spin
orientation $s=${\small$\uparrow,\!\downarrow$} and the layer
separation is zero.

In order to formally extract the effective CF-interaction
we consider the Lagrangian density of the two coupled Chern-Simons
liquids of charge $e$
\begin{equation}
{\cal L} ({\bf r},t) = {\cal L}_{\rm F} ({\bf r},t) + 
{\cal L}_{{\rm CS}} ({\bf r},t) + {\cal L}_{{\rm I}} ({\bf r},t)  
\label{eq:1}
\end{equation}
with the kinetic term of the Fermions
\begin{eqnarray}
{\cal L}_{\rm F}({\bf r},t)&=&\sum_{s=\uparrow,\downarrow}
\psi^{\dagger}_{s}({\bf r},t) 
\Big\{i\hbar\partial_{t}+e a_0^{s}({\bf r},t)\Big.\nonumber \\
-\frac{1}{2m}\Big[i\hbar\nabla&+&\frac{e}{c} 
\Big({\bf A}({\bf r})-{\bf a}^{s}({\bf r},t)\Big)\Big]^2\Big\} 
\psi_{s}({\bf r},t)\,,  
\label{eq:2}
\end{eqnarray}
the gauge term (${\bf \hat{z}}$ perpendicular to the 2D plane)
\begin{equation}
{\cal L}_{{\rm CS}}({\bf r},t)=-\frac{e}{\tilde{\varphi}\,\phi_{0}}
\sum_{s=\uparrow,\downarrow}a_0^{s}({\bf r},t)\, 
{\bf \hat{z}}\cdot\nabla\times{\bf a}^{s}
({\bf r},t)\,,
\end{equation}
and the contribution of the Coulomb interaction
\begin{equation}
{\cal L}_{{\rm I}}({\bf r},t)=-{\frac{1}{2}}\sum_{s,s'} \int d^2
r' \rho_{s}({\bf r},t) V({\bf r}-{\bf r}')
\rho_{s'} ({\bf r}',t).
 \label{lc}
\end{equation}
Here, $\rho_{s}({\bf r},t)\equiv \psi^{\dagger}_{s}({\bf r},t) \psi_{s}({\bf
  r},t)$ is the density of the Fermions with spin orientation $s$, ${\bf A}$
the vector potential of the external magnetic field, $(a_{0}^{s},{\bf a}^{s})$
the gauge field for spin orientation $s$, and $V({\bf r})=e^2/\epsilon r$ the
Coulomb interaction.

The Chern-Simons Lagrangian ${\cal L}_{{\rm CS}}$ is responsible for
the attachment of $\tilde{\varphi}$ flux quanta $\phi_{0}\equiv hc/e$
to each Fermion, as can be seen by minimizing the action with respect
to the $a_{0}^{s}$-gauge field. This gives the constraint ${\bf
  \hat{z}}\cdot\nabla\times{\bf a}^{s}({\bf r},t)
/{\tilde{\varphi}\phi_{0}} =\rho_{s}({\bf r},t).$ Here, the flux
attachment for the two species of Fermions has been performed
independently. We assume $\tilde{\varphi}=2$, such that the mean
fictitious magnetic field cancels the external one at half filling,
$\nu\equiv \rho\phi_{0}/2B=1/2$.

We use the transverse gauge, $\nabla\cdot{\bf
  a}^{s}=0$. Then, the Bosonic variables associated with the gauge
field fluctuations are the {\em transverse} components of their
Fourier transforms, $a_{1}^{s}({\bf
  q},\omega)\equiv {\bf \hat{z} }
\cdot\hat{\bf q}\times[{\bf a}^{s}({\bf
  q},\omega)-\langle{\bf a}^{s}({\bf q},\omega)\rangle]$.  By
introducing the mean gauge field into ${\cal L}_{\rm F}$ the external
field ${\bf A}$ is canceled. From the terms linear in the charge $e$
and the momentum $-i\hbar \nabla$, one easily can extract the form of
the vertices connecting two Fermions with one gauge field fluctuation
operator $a_{\mu}^{s}({\bf q},\omega)$ ($\mu=0,1$)
\begin{equation}
v^{s}_{\mu}({\bf k},{\bf k}+{\bf q})=\biggl(
\begin{array}{c}
e\\
\frac{\hbar e}{mc}\,{\bf \hat{z}}
\cdot\frac{{\bf k}\times{\bf q}}{|{\bf q}|}
\end{array}
\biggr).
\end{equation}
In addition, there is a Fermion-gauge field coupling term quadratic in
the fluctuation operators.

In order to derive the effective interaction between the CFs we
generalize the diagrammatic Chern-Simons Fermi liquid description of
FQHE states to include the spin. First, one inserts the
above relation between the charge density and the gauge field
into (\ref{lc}) and introduces the gauge field fluctuations such that
${\cal L}_{\rm CS}+{\cal L}_{\rm I}$ describes the free gauge field.
The effective CF interaction can then be obtained from the coupling
terms in ${\cal L}_{\rm F}({\bf r},t)$ by tracing out the gauge field
fluctuations. At imaginary time, one gets for the effective
interaction in the Euclidean action
\begin{eqnarray}
&V_{\mu\nu}^{s,s'}({\bf k},{\bf k'},{\bf q};\Omega_{n})
=v^{s}_{\mu}({\bf k},{\bf k}+{\bf q})\,v^{s'}_{\nu}({\bf 
k'},{\bf k'}-{\bf q})\nonumber\\
&\qquad\qquad\times[{\cal D}^{+}_{\mu\nu}({\bf q},\Omega_{n})+
(2\delta_{ss'}-1){\cal D}^{-}_{\mu\nu}({\bf q},\Omega_{n})].
\label{interaction}
\end{eqnarray}
This describes scattering of CFs from states with spin $s$, $s'$ and
momenta $\hbar{\bf k}$, and $\hbar{\bf k'}$ into states with
$\hbar({\bf k}+{\bf q})$, $\hbar({\bf k'}-{\bf q})$ by exchanging a
gauge field quantum with momentum $\hbar{\bf q}$ and frequency
$\Omega_{n}=2\pi nk_{\rm B}T/\hbar$ ($n$ integer).

The effective interaction contains the RPA gauge field propagators
${\cal D}^{\alpha}_{\mu\nu}({\bf q},\tau)\equiv (1/\hbar)\langle
{\bf T}_{\tau} a_{\mu}^{\alpha}({\bf q},\tau) a^{\alpha}_{\nu}(-{\bf
  q},0)\rangle$ (${\bf T}_{\tau}$ time ordering operator and $\alpha
=\pm $) evaluated in terms of the symmetric and antisymmetric
combinations of the gauge field fluctuations
$a^{\alpha}_{\mu}=(a^{\uparrow}_{\mu}+\alpha
a^{\downarrow}_{\mu})/{2}$.  In terms of the current-current
correlation functions for free Fermions at zero magnetic field,
$\Pi^{0}_{\nu}\equiv \Pi^{0}_{\nu\nu}({\bf q},\Omega_{n})$ one has
\begin{equation}
({\cal D}^{-1})^{\alpha}_{\mu\nu}({\bf q},\Omega_{n})=
 \left(
\begin{array}{cc}
-\Pi^{0}_{0}&
\qquad\frac{ieq}{\phi_{0}}\\
-\frac{ieq}{\phi_{0}}&\qquad
\frac{q^{2}V(q)}{\phi_{0}^{2}}\delta_{+,\alpha}-\Pi^{0}_{1}
\end{array}
\right)
\end{equation}

It can be shown that the dominant small-momentum small-energy contributions of
the above symmetric and antisymmetric propagators correspond to $\mu=\nu=1$
\cite{Bonesteel96}. For $\Omega_{n}\ll v_{{\rm F}}q\ll v_{{\rm F}} k_{{\rm
    F}}$, $\Pi^{0}_{0}\simeq e^{2}m/\pi\hbar^{2}$, $\Pi^{0}_{1}\simeq
-(e^{2}q^{2}/12\pi +2|\Omega_{n}|e^{2}\rho/v_{{\rm F} }q)/mc^{2}$, such that
\begin{equation}
{\cal D}^{\pm}_{11}({\bf q},\Omega_{n})\approx \frac{q}{\alpha_{\pm}\,
q^{(5\mp 1)/2}+\eta\, |\Omega_{n}|}
\end{equation}
with the constants $\eta=2e^{2}\rho/mc^{2}v_{{\rm F}}$,
$\alpha_{+}=e^{2}/\epsilon\phi_{0}^{2}$, $\alpha_{-}=4\pi\hbar^{2}/3
m\phi_{0}^{2}$. For small wave numbers, and frequency $\Omega_{n}\to 0$, the
antisymmetric propagator ${\cal D}^{-}_{11}({\bf q},\Omega_{n})$ dominates.
This does not contain the Coulomb interaction, as pointed out earlier
\cite{Bonesteel96}. In the Cooper channel, the effective interaction
$V_{11}^{s,-s}({\bf k},-{\bf k},{\bf q};\Omega_{n})$ is attractive for
$\Omega_{n} \to 0$ due to $D_{11}^{-}({\bf q},0)\propto q^{-2}$ ($q\to 0$) and
can lead to an instability towards the formation of spin-singlet pairs of CFs
in a single QHE layer.

We calculate the quasi-particle energy gap for the Cooper channel in mean
field approximation \cite{Khves93,Bonesteel96} using the Nambu field $
\Phi^{\dagger}_{{\bf k}}(\tau )=( c^{\dagger}_{\uparrow}({\bf k},\tau ),\,
c_{\downarrow}(-{\bf k},\tau )) $ where $c_{s}({\bf k},\tau), \,c_{s}^{\dagger
  }({\bf k},\tau)$ are the Fermion annihilation and creation operators,
respectively, for spin $s$ and momentum $\hbar {\bf k}$ at imaginary time
$\tau $. The Green function ${\cal G}({\bf k},\tau )=-(1/\hbar) \langle {\rm
  T}_{\tau }\Phi _{{\bf k}}(\tau )\Phi _{{\bf k}}^{\dagger }(0)\rangle $ is a
$2\times 2$ matrix, ${\cal G}_{ij}$, that is obtained by inverting the Dyson
equation
\begin{equation}
{\cal G}^{-1}({\bf k},\omega_{n})=
{\cal G}_{0}^{-1}({\bf k},\omega_{n})-\Sigma({\bf k},\omega_{n}).  
\label{Dyson}
\end{equation}
Here, ${\cal G}_{0}({\bf k},\omega_{n})=
[i\sigma_{0}\hbar\omega_{n}-\sigma_{3}(\hbar^{2}k^{2}/2m-\mu)]^{-1}$
is the (diagonal) Green function for a free Fermion pair. Here,
$\sigma_{0}$ is the unit matrix, $\sigma_{3}$ the Pauli matrix, $\mu$
the chemical potential, and $\omega _{n}=(2n+1)\pi k_{\rm B}T/\hbar$ a
Fermionic frequency. The self-energy $\Sigma ({\bf k},\omega_{n})$ is
a $2\times 2$-matrix $\Sigma_{ij}$ with the diagonal elements
describing the renormalization of the Fermion mass, while the
off-diagonal matrix element defines the pair-breaking gap $\Delta$
\begin{equation}
  \label{gap}
\Delta({\bf k},\omega
_{n})\equiv\frac{i\hbar\omega_{n}\Sigma_{12}({\bf k},\omega
_{n})}{i\hbar\omega_{n}-\Sigma_{11}({\bf k},\omega _{n})}.
\end{equation}

For the self-energy we use only the Fock term
\cite{Khves93,Nagaosa90},
\begin{eqnarray}
\Sigma_{ij}({\bf k},\omega _{n}) &=&
\frac{k_{\rm B}T}{(2\pi )^{2}}
\sum_{n'}\int d{\bf q}\,{\cal G}_{ij}({\bf k}-{\bf q},\omega_{n}-\omega _{n'})
\nonumber\\
&&\qquad\qquad\times\left[\delta_{ij}
V_{11}^{s,s}({\bf k},{\bf k},{\bf q};\omega _{n'})\right.\nonumber\\
&&+\left.(\delta_{ij}-1)
V_{11}^{s,-s}({\bf k},-{\bf k},{\bf q};\omega _{n'})\right]
\label{Selfen1}
\end{eqnarray}
In order to evaluate this explicitly, we assume ${\bf k}\approx {\bf k}_{\rm
  F}$. Carrying out the necessary integrations, by analytical continuation to
real frequencies, $\omega_{n}\to -i\omega+\delta$, using the spectral
representation of the Green function, and combining the self-energy equation
with the above (\ref{gap}) one obtains an implicit equation for
$\Delta(\omega)\equiv \Delta({\bf k}_{\rm F}, \omega)$ in the zero-temperature
limit,
\begin{eqnarray}
  \label{delta}
\Delta(\omega)&=&\frac{C}{\omega }\int_{-\infty}^{\infty}
{\rm d}z\,\Theta(\hbar|z|-\Delta(z))\nonumber\\
&&\times\frac{[zK_{1}(z)\Delta(\omega)-\omega\Delta(z)K_{2}(z)]}
{[\hbar^{2}z^{2}-\Delta^{2}(z)]^{1/2}}
\end{eqnarray}
with the constant $C=e^{2}\hbar k_{\rm F}/8\pi^{3}mc^{2}$ 
($E_{\rm F}$ Fermi energy),
\begin{equation}
  \label{eq:22}
 K_{j}(z)\equiv {\rm sgn}(z)[F_{\omega}^{-}(z)-(-1)^{j}F_{\omega}^{+}(z)]\,, 
\end{equation}
($j=1,2$) and
\begin{equation}
  \label{eq:20}
F_{\omega}^{\pm}(z)=\int_{-\infty}^{\infty}{\rm d}\zeta\,
\frac{{\rm sgn}(z)+{\rm sgn}(\zeta)}{\zeta+z-\omega-i\delta}\,
{\rm Im}D^{\pm}(\zeta)
\end{equation}
where $D^{\pm}=-\int_{0}^{\infty}{\rm d}q\,D^{\pm}_{11}(q,-i\zeta)$.
By evaluating the integrals one obtains for $\omega \to 0$ to leading order
the self-consistency condition
\begin{equation}
  \label{eq:21}
  1=C^{-}x^{1/3}-C^{+}[\ln{(\omega_{0}x/C^{+})}]^{2}\,.
\end{equation}
with $x=E_{\rm F}/\Delta$ and a cutoff parameter $\omega_{0}\gg 1$.  Apart
from the values of the constants, this result is similar to the one obtained
for the double-layer system \cite{Bonesteel96}. The first term, with the
numerical constant $C^{-}=\sqrt\pi(2/3)^{4/3}\Gamma(7/6)/\Gamma(2/3)\approx
1.4$, describes the contribution due to $D^{-}$. The second term, with the
prefactor $C^{+}=(1/2\pi k_{\rm F}\ell_{\rm C})(E_{\rm F}/E_{\rm C})$, is due
to $D^{+}$ and stems from the interaction between particles; $E_{\rm C}$ is
the Coulomb repulsion energy of two particles at distance $\ell_{\rm C}$.
Independent of the value of $C^{+}$, there is {\em always} a solution to this
equation. This does not contradict the results of \cite{bonesteel99} since we
considered here non-screened, long-range Coulomb interaction between the
particles. For $E_{\rm F}$ larger than $E_{\rm C}$, however, $\Delta$ becomes
vanishingly small. 

Equation (\ref{eq:21}) indicates that in a {\em single} QHE layer, when two
LLs intersect at the Fermi energy in a perpendicular magnetic field, the
system becomes unstable against formation of a spin-singlet state due to
attractive coupling of CFs via the gauge field fluctuations. The resulting
condensate state is similar to the macroscopic state induced in a
superconductor by the electron-phonon coupling.

Let us consider experimental consequences. First, the existence of the
spin-singlet condensate state contributes towards understanding the
extra-plateaus in the CF spin polarization experiments of \cite{Kukushkin}.
The splitting between CFLLs with spin up and spin down behaves as $\sqrt{B}$
for small $B$, and is proportional to $B$ for large $B$. Spin-up and spin-down
components of different CFLLs intersect. As an example, we consider $\nu\equiv
p/(2p+1)=2/5$. This corresponds to two filled CFLLs ($p=2$). We adjust the
Fermi level to the energy where the spin-down Zeeman level of the lowest CFLL
becomes degenerate with the spin-up Zeeman level of the first CFLL. For
magnetic fields smaller than the one corresponding to the point of degeneracy,
$B_{\rm c}$, only the Zeeman levels of the lowest CFLL are occupied at zero
temperature. The spin polarization vanishes,
$\gamma=(\rho_{\downarrow}-\rho_{\uparrow})/
(\rho_{\downarrow}+\rho_{\uparrow})=0$. Magnetic fields above $B_{\rm c}$
yield $\gamma=1$. Exactly at $B_{\rm c}$, two half-filled CFLLs can be formed
when defining the filling factor in terms of the ratio between the number of
CF and the number of ''effective'' flux quanta crossing the sample. In analogy
to the above, one can perform a gauge transformation leading to ''second
generation'' CFs. The corresponding gauge fluctuations mediate an effective
attractive interaction. This leads to the formation of a condensate. From the
experiments, one estimates $k_{\rm F}\ell_{\rm C}\approx 1$ and $E_{\rm
  F}/2\pi E_{\rm C}\approx 0.01$, such that $\Delta\approx 0.3 E_{\rm F}$. This
value is more or less consistent with the experimental observations concerning
the spin flip gap which give values of the order $0.2\,$K and the experimental
Fermi energy of 1$\,$K \cite{Kukushkin}, though this numerical estimate is
already somewhat outside the range of validity of the asymptotic condition
(\ref{eq:21}). The existence of the gap at the crossing point implies that in
a region of magnetic fields around this point, where the energy difference
between the CFLLs is less then $\Delta$, the condensate remains stable. The
formation of such a state of singlet CF-pairs was then responsible for the
formation of a plateau exactly at half the distance between the neighboring
plateaus in an interval of magnetic fields near the crossover point.

Second, one can suspect that mechanisms similar to the above can also lead to
instabilities at other filling factors. Near degeneracy, non-equal but
commensurate fillings of spin-up and spin-down states might yield collective
instabilities with intermediate spin polarizations. For instance, consider
$\nu=2/3$. This corresponds to two (Zeeman-split) CFLLs with $p=-2$. When
changing the magnetic field, the uppermost (spin-down) Zeeman level of the
lowest CFLL and the lower (spin-up) Zeeman level of the first CFLL intersect.
When the electron density is adjusted to the degeneracy point, the lowest
spin-up CFLL is fully occupied, while only half of the states in the
degenerated levels are filled. A priori, it is not clear how the Fermions are
distributed among these states. For the ground state all possible
configurations have to be taken into account. Above we have considered the
special case of both levels being half-filled. However, also other
configurations can yield stable states. Assume, for instance, that 3/4 of the
spin-up level of the first CFLL and 1/4 of the lowest spin-down CFLL are
occupied at degeneracy and that this, together with the totally occupied
lowest spin-up Zeeman level, would correspond to a stable state via the
effective interaction. This would give an average spin polarization of
$\gamma=3/4$. If such a state was stable in the presence of an in-plane
component of the magnetic field, it could account for the recently reported
3/4-state observed in NMR at $\nu =2/3$ \cite{f01}.

Third, the possibility of generating long-range spin-pairing correlations in a
single 2D Hall sample, similar to those discussed previously for QHE double
layers \cite{fw01,s00}, by tuning the density and the magnetic field to induce
the above crossing between spin-up and spin-down LL leads to interesting
speculations. For instance, consider two QHE systems in the same plane, say at
$\nu=2/5$, separated by a tunnel junction.  By tuning the two densities to the
value of the point of degeneracy a ''Josephson current'' should flow. Such a
current should vanish as soon as one of the two densities was detuned.

\vspace{-1mm}
In conclusion, we have considered two Landau levels with opposite
spins tuned to intersect at filling factor $1/2$ at the Fermi level.
By applying the Chern-Simons gauge transformation, we have derived an
effective attractive CF interaction. This yields an instability
towards a spin-singlet condensate. We have discussed several
experimental consequences. In order to observe the predicted
spin-singlet state, a close-to-zero in-plane component of the magnetic
field should be necessary as has been achieved in the
spin-polarization experiments done in the region of the FQHE. Our
results suggest that different occupations of spin-up and spin-down
LLs could account for instabilities at other fractional polarizations
and that an in-plane component of the magnetic field could account for
an anisotropic spin-singlet condensate.

\vspace{-0.7mm}
This work has been supported by the European Union via the TMR and RTN
programmes (FMRX-CT98-0180, HPRN-CT2000-00144), by the Deutsche
Forschungsgemeinschaft within the Schwerpunkt ``Quanten-Hall-Effekt'',
and by the Italian MURST via PRIN00.

\end{document}